\documentclass[aps,prb,twocolumn,groupedaddress,showpacs]{revtex4}
\usepackage{graphicx}
\usepackage{amsmath,amssymb,amsfonts}
\usepackage{natbib}

\newcommand{\bd}{\begin{displaymath}}
\newcommand{\ed}{\end{displaymath}}
\newcommand{\be}{\begin{equation}}
\newcommand{\ee}{\end{equation}}
\newcommand{\bs}{\begin{subequations}}
\newcommand{\es}{\end{subequations}}
\newcommand{\ba}{\begin{eqnarray}}
\newcommand{\ea}{\end{eqnarray}}

\begin{document}

\title{A causal look into the quantum Talbot effect}

\author{A. S. Sanz}
\email{asanz@imaff.cfmac.csic.es}

\author{S. Miret-Art\'es}
\email{s.miret@imaff.cfmac.csic.es}

\affiliation{Instituto de Matem\'aticas y F\'{\i}sica Fundamental,
Consejo Superior de Investigaciones Cient\'{\i}ficas,
Serrano 123, 28006 Madrid, Spain}

\date{\today}

\begin{abstract}
A well-known phenomenon in both optics and quantum mechanics is the
so-called Talbot effect.
This near field interference effect arises when infinitely periodic
diffracting structures or gratings are illuminated by highly coherent
light or particle beams.
Typical diffraction patterns known as quantum carpets are then
observed.
Here the authors provide an insightful picture of this nonlocal
phenomenon as well as its classical limit in terms of Bohmian
mechanics, also showing the causal reasons and conditions that
explain its appearance.
As an illustration, theoretical results obtained from diffraction of
thermal He atoms by both $N$-slit arrays and weak corrugated surfaces
are analyzed and discussed.
Moreover, the authors also explain in terms of what they call the
Talbot-Beeby effect how realistic interaction potentials induce shifts
and distortions in the corresponding quantum carpets.
\end{abstract}


\maketitle


\section{Introduction}
\label{sec:1}

Particle diffraction by different types of devices (slits, gratings,
or surfaces) has become a standard technique to test the validity of
quantum mechanics.
This is confirmed by a large amount of experiments, ranging from tiny
objects (e.g., electrons, neutrons, single atoms, or small clusters)
to more complex, mesoscopic-size systems (e.g., fullerenes, large
biomolecues, or Bose-Einstein condensates).
Experiments with large, structured particles induce to think that,
in principle, there are no size-dependent limits to observe particle
diffraction, except those associated with the preservation of the
system coherence.
This is a very important issue, since the system coherence may
disappear relatively fast as a consequence of the many eventual degrees
of freedom involved in the process, either related to the surrounding
environment (external) or associated with the own structure of the
particle (internal).

In 1836, when characterizing optical gratings, Talbot\cite{talbot}
observed a repetition of alternate color bands of complementary colors
(red-green and blue-yellow) at certain distances from the grating.
About 50 years later, in 1881, Rayleigh\cite{rayleigh} proved that this
phenomenon is a consequence of the diffraction of highly spatially
coherent (plane) waves by gratings; the color band structure observed
is a manifestation of the periodicity and the shape of the grating.
The alternation of complementary color bands occurs at integer
multiples of $z_T = d^2/\lambda$, the {\it Talbot distance}, where
$d$ and $\lambda$ are the grating period and the wavelength of the
incident plane wave, respectively; bands with equal color patterns
thus repeat at integer multiples of $2z_T$.
The Talbot effect has important technological applications in optics,
such as image processing and testing or production of optical elements.
Similarly, its quantum-mechanical counterpart is relevant in electron
optics, where it has many applications within electron microscopy.
Moreover, this effect has been observed experimentally with heavy
particles, such as Na atoms\cite{chapman} or Bose-Einstein
condensates.\cite{deng}

Like in optics, $N$-slit arrays can also be considered typical examples
of quantum gratings.
When such arrays are illuminated by continuous, coherent wave fronts,
a continuous quantum flow is observed behind the slits.
This flow displays a typical pattern called {\it quantum
carpet}\cite{berry1} with periodicity $d$ along the direction
parallel to the plane containing the slits ($x$) and $2z_T$
along the propagation direction ($z$).
Full {\it recurrences} along $z$ are found at integer multiples of
$z_T$; these recurrences are the direct analog of the color bands
observed when working with optical gratings.
A recurrence that coincides with the initial state describing the
(diffracted) system is called a {\it revival} of such a state and
appears at integer multiples of $z_T$; for even integers, the state
looks exactly the same as the initial one, while for odd integers
it is shifted half a period ($d/2$) with respect to the latter.
Recurrences at fractions of $z_T$ consist of superposed images of the
initial state with itself.
Indeed, if the boundary conditions of the slits are ``sharp'' (the
{\it window function} is not differentiable at the borders of the
slit), one can observe {\it fractal} structures at irrational
fractions of $z_T$, which give rise to {\it fractal
carpets}.\cite{berry1,wojcik,berry2,hall2,amanatidis}

\begin{figure}
 \includegraphics[width=8.2cm]{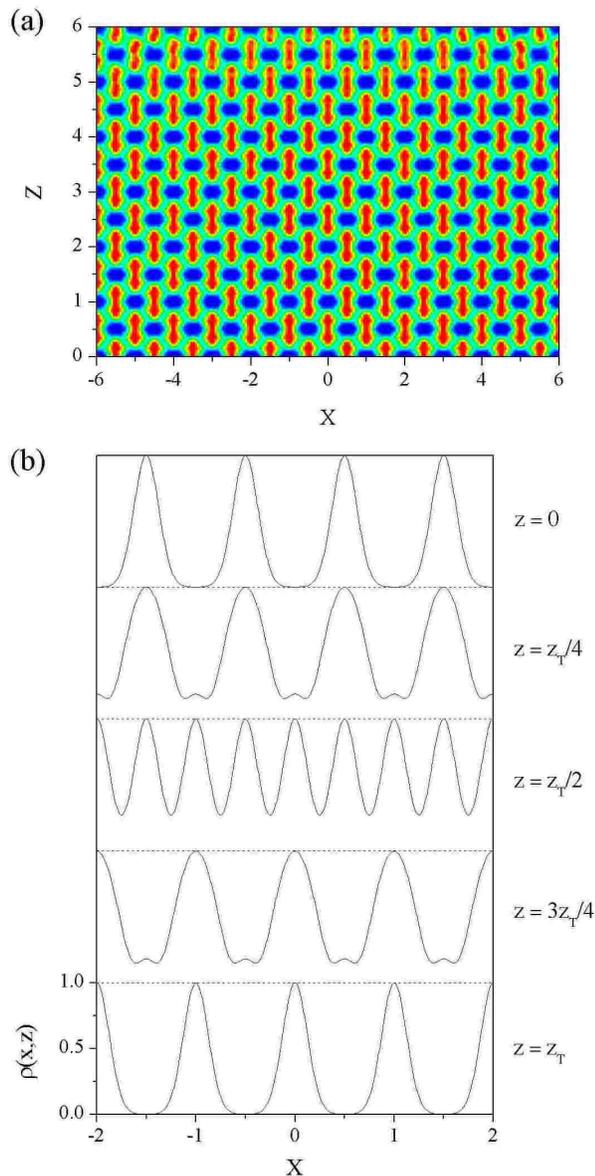}
 \caption{\label{fig1}
  (Color online) (a) Quantum carpet for He diffracted by a 50-slit
  array.
  The color scale, from blue to red, indicates increasing values of
  the probability density.
  (b) Snapshots of the probability density from $z = 0$ to $z = z_T$.
  In both panels, the $x$ distance is scaled in units of the grating
  period $d$ and $z$ in units of twice the Talbot distance ($2z_T =
  2d^2/\lambda$).}
\end{figure}

The Talbot carpet associated with the diffraction of a monochromatic
beam of He atoms ($E_z = 21$~meV, $\lambda = 2\pi\hbar/\sqrt{2mE_z}
= 0.991$~\AA) by a grating consisting of 50 Gaussian slits with period
$d = 3.6$~\AA\ is displayed in Fig.~\ref{fig1}(a).
Here, the $x$ variable is normalized to the grating period, $d$, and
the $z$ variable to twice the Talbot distance $2z_T$, with
$z_T = 13.08$~\AA\ (throughout this work we will follow the same
convention).
The carpet appears when we record $\rho_t(x,z) = |\Psi_t(x,z)|^2$
along $x$ at different times, monitored in terms of $z$ (the wave
propagates along the $z$ direction at a constant speed $v_z =
2\pi\hbar/m\lambda$; see below for details).
That is, at a given time $t_{\rm plot}$ we have plotted the ``slice''
$\rho_{t_{\rm plot}}(x,z_{\rm plot})$ of the probability density, where
$x$ runs over all possible values along a certain range (in the figure,
from $-$6 to 6) and the $z$ value is fixed: $z_{\rm plot} =
z(t_{\rm plot}) = v_z t_{\rm plot}$.
Some of these slices are shown individually in Fig.~\ref{fig1}(b),
from $z = 0$ to $z = z_T$.
As mentioned above, revivals of the initial state occur at $z_T$
(shifted half a period), while the other slices manifest vague
reminiscences of $\rho_0(x,0)$.
After reaching $z = z_T$, the diffracted beam will evolve towards
$2z_T$ following the inverse sequence to that shown
in Fig.~\ref{fig1}(b) [see Fig.~\ref{fig1}(a)], then an exact revival
of the initial probability distribution will be observed.
This process repeats endlessly unless the grating has a finite
size.
With real gratings its extension is limited to distances of the order
of the grating size.
At such distances the carpet gradually fades, and the typical
Fraunhofer patterns that characterize most particle diffraction
experiments start  becoming apparent.

Although Talbot patterns are well known in the literature, as far as
we know there is no detailed explanation for them in causal terms.
This work is aimed to provide a full causal interpretation for this
phenomenon as well as for its relationship with the Fraunhofer
diffraction and the classical limit.
For this purpose, we have chosen Bohmian mechanics, where the standard
wave picture is replaced by trajectories in configuration space.
The use of such trajectories is very important in order to shed some
light into real experiments, where a picture in terms of the motion of
individual particles is always highly desirable (specially if particles
follow the quantum flow, unlike other classical representations).
This is an important issue, for instance, in surface science
experiments, where the illumination of a surface for its study
and characterization is limited in extension.
Note that, when assuming that no imperfections are present, surfaces
can be considered as gratings analogous to $N$-slit arrays because of
their periodicity.
Nonetheless, the short-range attractive interaction undergone by the
diffracted particles near the surface leads to a slight shift and
distortion of the Talbot period.
We have called this effect the {\it Talbot-Beeby effect} since the
variation of the diffracted beam wavelength due to presence of
attractive wells is known as the {\it Beeby correction}\cite{beeby}
in surface scattering.

This work is organized as follows.
In order to be self-contained, a rigorous mathematical approach in
terms of standard quantum mechanics will be presented in
Sec.~\ref{sec:2}, analyzing both the appearance of the Talbot effect in
gratings of Gaussian slits and the transition to the Fraunhofer regime.
In Sec.~\ref{sec:3} we describe the application of Bohmian mechanics
to the problems discussed in the previous section.
A discussion of the Talbot and Talbot-Beeby effects in diffraction
of He atoms by both an $N$-slit array and the Cu(110) surface is
given in Secs.~\ref{sec:4} and \ref{sec:5}, respectively.
In the light of these results, the meaning of the classical limit and
the quantum-classical correspondence is discussed in Sec.~\ref{sec:6}.
Finally, the main conclusions from this work are summarized
in Sec.~\ref{sec:7}.


\section{Wave approach to the Talbot effect}
\label{sec:2}

Rigorous analytical studies of the Talbot effect from an optical
viewpoint can be found, for instance, in Ref.~\onlinecite{winthrop}.
Here we provide an alternative quantum-mechanical derivation
highlighting the physical aspects underlying this phenomenon, in
particular, (i) the role of the superposition principle and (ii) the
analogy/equivalence between Talbot patterns and the carpets observed
in multimode cavities (e.g., waveguides).
For simplicity, we focus on gratings constituted by Gaussian
slits,\cite{sanz2,sanz3} i.e., characterized by Gaussian transmission
functions, though our analysis can be straightforwardly generalized
to any kind of periodic grating.
Gaussian transmissions can be observed, for instance, when studying the
diffraction of a monochromatic beam (with wavelength  $\lambda$) by a
soft, repulsive (exponential) potential barrier with an infinity of
identical holes (slits).\cite{sanz3}


\subsubsection{Single Gaussian slit}

To understand the physics associated with the time evolution of the
diffracted wave function by an infinite Gaussian grating, it is
important to consider first the dynamics of a single diffracted
Gaussian function.
Assuming the initial time ($t=0$) as the instant when the incident
wave has just passed through the slit with perpendicular velocity
($v_x = 0$), the initial state can be expressed as
\be
 \Phi_0(x,z) =
  A(0) \ \! e^{- x^2/4\sigma_x^2 - z^2/4\sigma_z^2 + i p_z z/\hbar} ,
 \label{eqn1}
\ee
where $A(0) = (2\pi\sigma_x\sigma_z)^{-1/2}$ is the norm and
$\sigma_x$ and $\sigma_z$ the initial widths in the $x$ and $z$
directions, respectively.
Within this description, note that $\sigma_x$ is related to the
transmission function of the slit, giving an idea of the aperture
dimension (i.e., the slit width); on the other hand, $\sigma_z$ is
an indicator of the monochromaticity of the incident beam (the limit
$\sigma_z \to \infty$ represents the case of a fully monochromatic
incident wave).

The time propagation of $\Phi_0$ is given by
\begin{multline}
 \Phi_t(x,z) = A(t) \ \! e^{- x^2/4\tilde{\sigma}_{x,t}\sigma_x
   - (z - z_t)^2/4\tilde{\sigma}_{z,t}\sigma_z} \\
  \times \! e^{i p_z (z - z_t)/\hbar + i E_z t/\hbar} ,
 \label{eqn2}
\end{multline}
%
with $A(t) = (2\pi\tilde{\sigma}_{x,t}\tilde{\sigma}_{z,t})^{-1/2}$,
$z_t = v_z t$, $E_z = p_z^2/2m$, and
\be
 \tilde{\sigma}_{i,t} =
  \sigma_i \left( 1 + \frac{i\hbar t}{2m\sigma_i^2} \right) ,
  \qquad i = x, z ,
 \label{eqn3}
\ee
where
\be
 \sigma_{i,t} = |\tilde{\sigma}_{i,t}| =
  \sigma_i \sqrt{1 + \frac{\hbar^2 t^2}{4m^2 \sigma_i^4}} ,
  \qquad i = x, z
 \label{eqn4}
\ee
gives the instantaneous width along each direction.
From Eq.~(\ref{eqn4}) a time scale $t_s = 2m\sigma_i^2/\hbar$
separating two regimes with different ``spreading rates'' can
be defined.
If $t \ll t_s$, the Gaussian widths almost remain the same
($\sigma_{i,t} \approx \sigma_i$).
On the other hand, if $t \gg t_s$, the Gaussian widths undergo a
linear increase with time ($\sigma_{i,t} \approx \hbar t/2m\sigma_i$).
In this way, choosing $\sigma_x = d/8$ and $\sigma_z = d$, $\Phi_t$
will not almost display any increase in its size along $z$, but only
along $x$.
Accordingly, Eq.~(\ref{eqn2}) can be reexpressed as
\begin{multline}
 \Phi_t(x,z) = A'(t) \ \! e^{- x^2/4\tilde{\sigma}_{x,t}\sigma_x
    - (z - z_t)^2/4\sigma_z^2} \\
   \times \ \! e^{i p_z (z - z_t)/\hbar + i E_z t/\hbar} ,
 \label{eqn5}
\end{multline}
%
where $A'(t) = (2\pi\tilde{\sigma}_{x,t}\sigma_z)^{-1/2}$.
Note that this will ensure the overlapping of Gaussian functions
(along the $x$ direction) in the case of the grating, and therefore
the appearance of interference.

For hard slits, the interaction potential is zero everywhere except
along the grating ($z = 0$), where it is infinity.
Thus, there is no coupling between both coordinates, $x$ and $z$, and
the time evolution of the wave function along each direction can be
studied separately.
Since the interference takes place along $x$, we will only focus
on the part of the wave function depending on this variable,
\be
 \phi_t (x) = A_x(t) \ \! e^{- x^2/4\tilde{\sigma}_{x,t}\sigma_x} ,
 \label{eqn6}
\ee
where $A_x(t) = (2\pi\tilde{\sigma}_{x,t}^2)^{-1/4}$.
Regarding the component of the wave function along $z$, it is only
important to know that it propagates at a constant speed $v_z$, i.e.,
its centroid evolves according to $z(t) = v_z t$.
This allows to simplify our analysis, reducing it to a one-dimensional
time-dependent problem, where the $z$ coordinate contains the same
information as $t$.
Thus, from now on the subscript $x$ will be dropped from the magnitudes
related to the spreading along the $x$ direction ($\sigma$ and $A$
instead of $\sigma_x$ and $A_x$, respectively), and from the momentum
associated with the $x$ coordinate ($p$ instead of $p_x$).


\subsubsection{Infinite periodic slit gratings}

According to Bloch's theorem,\cite{ashcroft} the problem of finding
the diffracted wave function associated with a given infinite periodic
potential reduces to determining the wave function associated with a
single {\it unit cell} of such a potential; the full wave function is
just a repetition of the latter, which satisfies the following
Born--von Karman boundary conditions:
\be
 \psi_t(x+d) = \psi_t(x) .
 \label{eqg2}
\ee
Thus, we assume that the total number of Gaussian slits amounts to
$N=2K +1$, the slits being centered at $x_0^{(k)} = k d$ (with
$k = 0$, $\pm 1$, $\pm 2$, \ldots, $\pm K$) and $z_0^{(k)} = 0$.
Our problem then reduces to only consider one of the Gaussian wave
packets $\phi_t$, constituting the full wave function.
We can choose, for instance, $\phi_t$ to be the wave packet with
$k = 0$, which is confined within the space region $x = \pm d/2$
and whose time evolution is described by Eq.~(\ref{eqn6}).
In direct analogy to surface science, from now on we will call the
space region of length $d$ enclosing each (initial) Gaussian wave
packet [i.e., $(k-1)d/2 \leq x \leq kd/2$] a unit cell.

As is well known, any unbound wave function can be represented as a
superposition of plane waves,
\be
 \phi_t (x) = \frac{1}{\sqrt{2\pi\hbar}}
  \int a(p) \ \! e^{ipx/\hbar - i\omega t} \ \! dp ,
\ee
with
\be
 a(p) = \frac{1}{\sqrt{2\pi\hbar}}
  \int \phi_0 (x) \ \! e^{-ipx/\hbar} \ \! dx .
\ee
Thus, Eq.~(\ref{eqn6}) can be expressed as
\be
 \phi_t(x) = \frac{(8\pi\sigma^2)^{1/4}}{2\pi\hbar}
  \int e^{- \sigma^2 p^2/\hbar^2 + ipx/\hbar - i\omega t} \ \! dp .
 \label{eqg4}
\ee
Because of the periodic boundary condition (\ref{eqg2}), not all
momenta are allowed; we then pass from a continuous basis of momenta
to a discrete one, and Eq.~(\ref{eqg4}) becomes
\be
 \phi_t (x) = \sqrt{\frac{1}{d}}
  \left( \frac{8\pi\sigma^2}{d^2} \right)^{1/4}
  \sum_{|n|=1}^\infty e^{- \sigma^2 p_n^2/\hbar^2 + ip_n x/\hbar
  - i \omega_n t} ,
 \label{eqg5}
\ee
where $p_n = 2\pi\hbar n/d$ and $\omega_n = 2\pi^2\hbar n^2/md^2$.

The initial (full) wave function describing the system (along the
$x$ direction) is assumed to be a coherent, nonoverlapping
superposition of identical Gaussian wave functions that propagate
along the $z$ direction with constant velocity,
\be
 \psi_0 (x) \propto
  \lim_{K\to\infty} A(0) \sum_{k=-K}^K e^{- (x - kd)^2/4\sigma^2} .
 \label{eqg1}
\ee
The corresponding time-evolved wave function is straightforwardly
obtained by replacing the Gaussian wave packets in Eq.~(\ref{eqg1})
by their time-dependent counterparts given by Eq.~(\ref{eqg5}).
This leads to
\be
 \psi_t (x) \propto \lim_{K\to\infty} (2K+1) \ \! \phi_t (x)
 \label{eqg6}
\ee
after rearranging terms.
As one would expect, Eq.~(\ref{eqg6}) indicates that all the
information regarding the infinite grating is contained within a
single unit cell, with the factor $2K+1$ arising from the total
number of unit cells considered.
To avoid the divergence introduced by this factor, we reexpress
Eq.~(\ref{eqg6}) as
\be
 \psi_t (x) = \lim_{K\to\infty} \frac{A(t)}{2K+1}
   \sum_{k=-K}^K e^{- (x - kd)^2/4\tilde{\sigma}_t\sigma} .
 \label{eqg1bt}
\ee

The smallest time elapsed necessary to observe a recurrence of the
wave function is determined by the condition
\be
 \psi_{t+\tau_r} (x) = \psi_t (x) .
 \label{eqg7}
\ee
Using Eq.~(\ref{eqg5}), it is straightforward to see that this
condition is satisfied whenever $e^{-i\omega_n\tau_r} = e^{-2\pi i}$
for all $n$, i.e., when $\tau_r = 2\pi/\omega_1 = md^2/\pi\hbar$.
At this time, the $z$ distance between two consecutive recurrences of
$\phi_t$ will be $z_r = v_z \tau_r = 2d^2/\lambda$, i.e., twice
the Talbot distance ($2z_T$).
As shown in Fig.~\ref{fig1}(b), there are also recurrences with
periodicity $z_r = d^2/\lambda = z_T$.
These recurrences result from considering the symmetry
\be
 \psi_{t+\tau_r/2} (x+d/2) = \psi_t (x) ,
 \label{eqg8}
\ee
which appears when taking the terms $e^{ip_n x/\hbar-i\omega_n\tau_r}$
as a whole.

\begin{figure}
 \includegraphics[width=7.8cm]{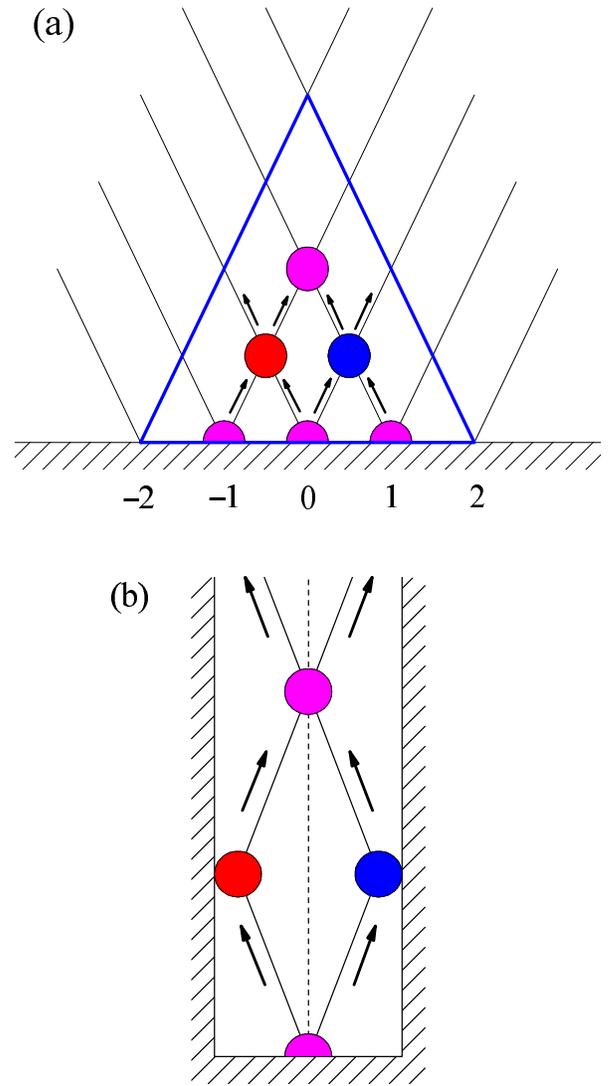}
 \caption{\label{fig2}
  (Color online) Schematic picture of the formation of revivals in
  (a) gratings of Gaussian slits and (b) multimode cavities.
  Arrows indicate the direction of the motion of the ``interfering''
  disks (see text for details) and straight lines represent the
  corresponding paths.
  In part (a), the numbers label different slits, and the blue line
  limits the Talbot region.}
\end{figure}

The observation of revivals can be explained by means of a simple
schematic picture, which is displayed in Fig.~\ref{fig2}(a).
To understand this picture we consider that each wave packet can be
represented by two ``interfering'' hard disks that propagate along
the $z$ direction with opposite $x$ velocities (the $x$ motion of
the center of mass is null, as expected from the propagation of a
real wave packet, which only advances along $z$).
These disks thus indicate the spreading of the wave packet.
Following the motion of the disks associated with $k = 0$, we observe
that they meet at $x = \pm d/2$ and $z = z_T$ with those arriving
from the neighboring slits ($k = \pm 1$).
Since all disks are identical, there will be full constructive
``interference'', this leading to the first revival.
The next revival will occur at $z = 2z_T$ and is caused by the
interference of the disks coming from $k = \pm 1$, but not from
$k = 0$; the disks leaving the slit with $k = 0$ will contribute
to the revivals observed at $x = \pm d$ and $z = 2z_T$.

From Fig.~\ref{fig2}(a) we can also infer an effective size for
Talbot structures when the grating periodicity is limited.
In the picture we have represented five slits; thus, after the
ingoing-moving disks corresponding to $k = \pm 2$ have interfered,
no Talbot revivals will be observed.
From the point of view of Gaussian wave packets, this will happen
after the size of those corresponding to the outermost slits is large
enough as to make them to interfere, i.e., when $2\sigma_t \approx
2Kd$.
Assuming that at that stage the width of the Gaussian functions
increases linearly with time, i.e., $\sigma_t \approx \hbar t/2m\sigma$,
the maximum time at which a (partial) revival can be observed is
$t_{\rm max} \approx 2Kdm\sigma/\hbar$.
Since the propagation along $z$ is also linear with time (at the speed
$v_z$), the $z$ distance where such a revival will be observable is
\be
 z_{\rm max} \approx
  v_z t_{\rm max} \approx 2z_T \ \! \frac{\pi (N-1)}{8}
 \label{eqzmax}
\ee
(we have particularized this expression to our case, where
$\sigma = d/8$).
Examples of the validity of this relation will be seen in
Sec.~\ref{sec:4}.
Beyond $z_{\rm max}$ a transition towards the Fraunhofer diffraction
regime, with its characteristic fringe patterns, starts to be observed.
This transition takes place beyond the boundaries of the Talbot region,
enclosed in Fig.~\ref{fig2}(a) by a blue triangle with height
$z_{\rm max}$ and basis $(N-1)d$.

As soon as the wave function (or part of it) leaves the Talbot region,
Bloch's theorem (and therefore the Born--von Karman boundary conditions)
is no longer applicable.
To describe the Fraunhofer diffraction, one has to start from
Eq.~(\ref{eqg1bt}), but having in mind that the grating has a
finite size.
Thus, at a relatively long $z$ distance from the grating, its
dimensions are negligible when compared with the $x$ distances
involved in the diffraction process [i.e., $x \gg (N-1)d$], and
therefore
\be
 e^{-(x-kd)^2/4\tilde{\sigma}_t\sigma} \approx
  e^{-\sigma^2 \kappa^2 x^2/z^2} \ \! e^{i \kappa x^2/2z} \ \!
  e^{-ik\kappa d x/z} ,
 \label{eqap}
\ee
where $\kappa = 2\pi/\lambda$.
Substituting the right hand side of this expression into
Eq.~(\ref{eqg1bt}) we obtain
\be
 \psi_t (x) \approx \frac{A(t)}{2K+1} \ \!
  e^{-\sigma^2 \kappa^2 x^2/z^2} \ \! e^{i \kappa x^2/2z}
  \sum_{k=-K}^K e^{- i k \kappa dx/z} .
 \label{eqg12}
\ee
Now, using the so-called paraxial approximation from
optics\cite{bornwolf} and considering $x/z = \tan \theta \approx
\sin \theta$ ($\theta$ is the observation angle), the probability
density (diffraction pattern) along $x$ can be expressed as
\be
 \varrho_t (x) = |A(t)|^2 \ \! e^{-\sigma_x^2 \kappa^2 x^2/z^2}
  \left[ \frac{\sin \left( N\kappa d\sin \theta/2 \right)}
  {N \sin \left( \kappa d\sin \theta/2 \right)} \right]^2 .
 \label{eqg13b}
\ee
In this expression, the term between square brackets is the
{\it structure factor}, which accounts for the interference
among the different diffracted (Gaussian) beams.
On the other hand, the normalized exponential is the {\it form factor},
which is related to the diffraction by a single unit cell.
Because of the information provided by these factors, they are
very useful to characterize optical grids\cite{bornwolf} as well as
periodic surfaces.\cite{ashcroft}

From the structure factor we note that the Fraunhofer fringes appear
in accordance to the diffraction (quantization) condition
\be
 \sin \theta = \ell \ \! \frac{\lambda}{d} , \qquad
  \ell = 0, \pm 1, \pm 2, \ldots
 \label{eqg14}
\ee
where $\ell$ is the {\it diffraction order}.
On the other hand, there will be a series of minima whenever
\be
 \sin \theta = \frac{\ell'}{N} \ \! \frac{\lambda}{d} ,
  \qquad |\ell' - \ell| = 1, 2, \ldots , N-1 .
 \label{eqg15}
\ee
As seen, there are $N$$-$1 minima between two consecutive principal
maxima and, consequently, $N$$-$2 secondary maxima; the height of
the latter is smaller than that of the principal maxima and
decreases\cite{sanz2} very fast with $N$.

Finally, a comment regarding the local spreading of each Gaussian wave
packet and the boundary condition Eq.~(\ref{eqg2}) is worth stressing.
According to Eq.~(\ref{eqn6}), after a certain time the size of every
Gaussian in (\ref{eqg1bt}) at $t=0$ will be such that it will extend
beyond the boundaries of its corresponding unit cell.
That is, the spreading of $\phi_t$, which is a local property, gives
rise to the appearance of a nonlocal behavior, where each part of the
resulting wave function, $\psi_t$, is strongly influenced by the
presence of the remaining ones.
Note, indeed, that the Talbot effect is precisely a nonlocal effect:
it emerges as a consequence of the overlapping of many identical wave
packets.
As will be seen below, Bohmian mechanics provides a natural picture
to this problem in terms of the regimes of motion associated with two
well-defined dynamic equilibria.


\subsubsection{Multimode cavities}

The Talbot effect is closely related to multimode
interference,\cite{kaplan1,kaplan2,nest} the interference process that
takes place when an infinity of modes of a cavity are superposed.
Waveguides are a typical example of multimode cavity where the wave
motion is constrained along one direction (e.g., $x$) and unbound along
the other (e.g., $z$).
Assuming no coupling between the motion along each direction, and that
the cavity is a square box along $x$, centered at $x = 0$ and with
length $d$, the time evolution of a Gaussian wave packet can be
expressed at any time in terms of the modes corresponding to this
cavity as
\begin{multline}
 \phi_t (x) = \sqrt{\frac{8}{d}}
  \left( \frac{2\pi\sigma^2}{d^2} \right)^{1/4} \\
  \times \sum_{n=0}^\infty e^{- \sigma^2 p_n^2/\hbar^2
  - i E_n t/\hbar} \cos (p_n x/\hbar) , \quad n = 0, 1, 2, \ldots
 \label{eqi2}
\end{multline}
%
with $p_n = (2n+1)\pi\hbar/d$ and $E_n = p_n^2/2m$.
It is easy to show\cite{sanz1} that recurrences in the probability
density arising from Eq.~(\ref{eqi2}) occur at integer multiples of
the period associated with the smallest frequency.
In our case, this frequency is $\omega_{1,0} = (E_1 - E_0)/ \hbar
= 4\pi^2\hbar/md^2$, and the period associated is
\be
 \tau_r = \frac{2\pi}{\omega_{1,0}} = \frac{md^2}{2\pi\hbar} .
 \label{eqi4}
\ee
We have to mention that the same periodicity can also be observed
in the wave function, except for a constant phase factor
[$\phi_{t+\tau_r} (x) = e^{i\varphi} \phi_t (x)$].

Since the wave function evolves at a constat speed along $z$, the
recurrences will also give rise to the formation of quantum carpets;
some examples of these carpets can be found, for instance, in
Ref.~\onlinecite{sanz2}.
These recurrences appear at integer multiples of the distance
$z_r = v_z \tau_r = d^2/\lambda$, which corresponds to the Talbot
distance obtained in the case of a periodic grating.
This is the analogous case to the revivals observed
at $z = 2z_T$ in grating systems.
To understand why the same type of recurrences cannot be observed in
both periodic gratings and cavities, it is very instructive to look at
the scheme represented in Fig.~\ref{fig2}(b).
This picture is equivalent to that shown in Fig.~\ref{fig2}(a), but
with the difference that the only possibility for the disks to cross is
when they meet again; no interference at $x = \pm d/2$ is possible.
In other words, while the quantization leading to Eq.~(\ref{eqg5})
arises from a ``matching'' condition at the borders of neighboring unit
cells (thus allowing interference), in Eq.~(\ref{eqi2}) it comes from
having impenetrable boundaries.


\section{Bohmian approach to the Talbot effect}
\label{sec:3}

The fundamental equations of Bohmian mechanics are commonly
derived by expressing the system wave function in polar
form,\cite{bohm,holland,wyatt}
\begin{equation}
 \Psi_t (x,z) =
  \rho_t^{1/2}(x,z) \ \! {\rm e}^{{\rm i} S_t (x,z)/\hbar} .
 \label{eq12}
\end{equation}
Here, $\rho_t$ is the probability density and $S_t$ is the
(real-valued) phase.
Substituting Eq.~(\ref{eq12}) into the time-dependent Schr\"odinger
equation and making a bit of algebra, one reaches two (real-valued)
coupled differential equations
\begin{eqnarray}
 & \displaystyle \frac{\partial \rho_t}{\partial t} + \nabla \cdot
   \displaystyle \left(\rho_t \ \frac{\nabla S_t}{m} \right)=0 , &
  \label{eq13} \\
 & \displaystyle \frac{\partial S_t}{\partial t}
   + \frac{(\nabla S_t)^2}{2m} + V + Q_t = 0 .
 & \label{eq14}
\end{eqnarray}
Equation (\ref{eq13}) is a continuity equation that ensures the
conservation of the quantum particle flux.
On the other hand, Eq.~(\ref{eq14}), more interesting from a dynamical
viewpoint, is a quantum Hamilton-Jacobi equation governing the particle
motion under the action of a total effective potential $V_t^{\rm eff}
= V + Q_t$.
The last term in the left hand side of this equation is the so-called
{\it quantum potential}
\begin{equation}
 Q_t =
  - \frac{\hbar^2}{2m}\frac{\nabla^2 \rho_t^{1/2}}{\rho_t^{1/2}} .
 \label{eq15}
\end{equation}
This context-dependent, nonlocal potential determines together with
$V$ the total force acting over the system.

In the classical Hamilton-Jacobi theory ($Q_t=0$), $S_t$ represents the action
of the system at a time $t$, and the trajectories describing the
evolution of the system correspond to the paths perpendicular to the
constant-action surfaces at each time.
Similarly, since the Schr\"odinger equation can be rewritten in terms
of the Hamilton-Jacobi equation (\ref{eq14}), $S_t$ can be interpreted
as a quantum action satisfying similar mathematical requirements as its
classical homologous.
In Bohmian mechanics the classical-like concept of trajectory thus
emerges in a natural manner; particle trajectories are defined as the
solutions of the equation of motion
\begin{equation}
 \dot{\bf r} = \frac{\nabla S_t}{m} = \frac{\hbar}{m} \
  {\rm Im} \left[ \Psi_t^{-1} \nabla \Psi_t \right] .
 \label{eq16}
\end{equation}
Moreover, since in Bohmian mechanics the system consists of a wave and
a particle, it is not necessary to specify the initial momentum for
each particle (as happens in classical mechanics), but only its initial
position, ${\bf r}_0 = (x_0,z_0)$, as well as the initial wave
function $\Psi_0$.
The initial momentum field is predetermined by $\Psi_0$ via
Eq.~(\ref{eq16}), and the statistical predictions of the standard
quantum mechanics are reproduced by considering an ensemble of
(noninteracting \cite{note2}) particles distributed according
to the initial probability density $\rho_0 = |\Psi_0|^2$.

Equation (\ref{eq16}) is well defined provided the wave function is
continuous and differentiable.
This is not the case, however, for quantum
fractals,\cite{wojcik,berry2,hall2,amanatidis} where the Bohmian
mechanics based on Eq.~(\ref{eq16}) is unable to offer a trajectory
picture for the corresponding wave functions.
This apparent incompleteness can be nevertheless ``bridged'' by taking
into account the decomposition of the quantum fractal as a sum of
(differentiable) eigenvectors of the Hamiltonian, and then redefining
Eq.~(\ref{eq16}) conveniently.
This is seen in detail in Ref.~\onlinecite{sanz1}, where a
generalization of the standard formulation of Bohmian mechanics
to include quantum fractals is given.

Prior to any calculation, some physical insight into slit diffraction
problems is possible by studying the properties of the velocity field.
Indeed, the fact that Bloch's theorem (together with the Born--von
Karman boundary conditions) holds simplifies this study from both
a conceptual and a computational perspective.
Conceptually, because the study of the full system reduces to only
understanding the dynamics within a single unit cell.
In this sense, the analysis is similar to that of having a multimode
cavity, as said above.
This implies a computational advantage: it allows to perform
calculations taking into account only what happens inside a single
unit cell and periodic boundary conditions [i.e., $\psi_t(x-d/2) =
\psi_t(x+d/2)$], thus reducing the computation efforts.
Of course, this simplification is only possible if we are under the
assumption of infinite periodic gratings or, at least, we are working
well inside the Talbot area, delimited by the blue triangle shown in
Fig.~\ref{fig2}(a).
Otherwise, this advantageous framework is no longer valid and the
full system wave function has to be considered.

According to the previous statements, substituting Eq.~(\ref{eqg5})
into (\ref{eq16}) for the $x$ coordinate yields
\begin{widetext}
\be
 \dot{x} = \frac{1}{m}
  \frac{\sum_{i,j} p_i \ \! e^{- \sigma^2 (p_i^2+p_j^2)/\hbar^2}
   \cos [(p_i - p_j)x/\hbar - (\omega_i - \omega_j)t]}
  {\sum_{i,j} e^{- \sigma^2 (p_i^2+p_j^2)/\hbar^2}
  \cos [(p_i - p_j)x/\hbar - (\omega_i - \omega_j)t]} ,
   \qquad i, j = \pm 1, \pm 2, \ldots
 \label{eqfull}
\ee
\end{widetext}
From this equation we can extract relevant information about the
physical properties of the quantum trajectories (or, equivalently,
their topology).
Note that the velocity field (\ref{eqfull}) satisfies exactly the same
symmetry conditions expressed by Eqs.~(\ref{eqg7}) and (\ref{eqg8}).
The particle motion is thus oscillatory, with the recurrences displayed
by $\rho_t$ occurring in those space regions where trajectories
accumulate.
However, unlike the disks depicted in Fig.~\ref{fig2}(a), the
trajectories associated with each single unit cell will remain inside
it at any time and will never cross those coming from other unit cells:
the particle motion is {\it bound}.
This fact is a manifestation of the {\it noncrossing property} of
Bohmian mechanics: trajectories can never pass through the same point
on configuration space at the same time due to the single valuedness
of the momentum field.
This explains why the physics along each unit cell (or, by extension,
the physics associated with infinite gratings) is so close to that
observed in multimode cavities.
The boundary periodic conditions give rise to the presence of
nonphysical {\it impenetrable} walls at $x = \pm d/2$, where the
velocity field (\ref{eqfull}) vanishes and therefore the corresponding
quantum trajectories will be just straight lines.
Due to the noncrossing property, these trajectories will act as
(impenetrable) boundaries for particles coming from different
neighboring slits.
This is the Bohmian causal explanation for the Born--von Karman periodic
boundary conditions.
The same effect is also found for the trajectories starting at the
center of the slits ($x = 0$), since the velocity field is also
zero along this symmetry line.
These trajectories evolve along two directions which
are specular to one another (with respect to $x = 0$).
Note that this goes beyond the classical-like picture provided by the
disks associated with a multimode cavity in Fig.~\ref{fig2}(b) in the
sense that it adds a constraint of different nature to the types of
motion that one can observe in quantum mechanics.
Moreover, this behavior is regardless of the motion along the
$z$ direction (provided there is no coupling between both
directions), where the corresponding equation of motion renders
\be
 z(t) = z_0 + v_z t .
\ee
That is, all particles display the same uniform rectilinear motion,
as happens in classical mechanics, since the swarm of particles is
basically guided by a plane wave.

If a size-limited grating is considered, after some time the
trajectories will be out of the Talbot region, and therefore the
description given above will no longer be applicable.
Far beyond the grating, one can appeal to the Fraunhofer approximation
in order to gain some insight on the topology of the trajectories.
In such a case, introducing Eq.~(\ref{eqg12}) into (\ref{eq16})
leads to
\be
 \dot{x} \approx \frac{\hbar \kappa}{m} \frac{x}{z}
  = v_z \ \! \frac{x}{z} .
 \label{eqb1}
\ee
Assuming that the probability density is only significant along the
quantized values of $x/z \approx \sin \theta$, where one observes
the Fraunhofer principal maxima (we neglect the presence of secondary
maxima), Eq.~(\ref{eqb1}) becomes
\be
 \dot{x} \approx \ell v_z \ \! \frac{\lambda}{d} .
\ee
This assumption is equivalent to considering that the diffracted wave
function consists of different independent plane waves, each one
characterized by a quantized momentum $p_\ell = \ell (2\pi\hbar/d)$.
This is a very important result that can be understood in terms of
two different equilibrium regimes.
The first equilibrium regime occurs in the Talbot/Fresnel region and
is characterized by what we could call an {\it equilibrium of momenta}.
That is, in this region the possible momenta satisfying a certain
quantization condition are selected.
This selection depends on the features defining the unit cell of the
grating (or, in other words, the interaction potential between the
grating and the diffracted particles).
The second equilibrium regime, which we could call {\it equilibrium
of configuration}, happens far beyond the grating, once the particle
distribution remains with the same shape (regardless of spreading
effects).
The transition from the momentum to the configuration equilibrium
is a direct consequence of the redistribution of momenta among the
different particles contributing to $\rho_t$.
These momenta make the swarm of particles to evolve in such a manner
that, at a certain distance from the grating, it will separate in
different beams, each moving with a different (average) momentum
$p_\ell = \ell (2\pi\hbar/d)$.
The equation of motion for the particles will then be
\be
 x(t) \approx x_0 + \frac{2\pi\hbar\ell}{md} \ \! t ,
\ee
where $\ell = 0$ denotes the classical direction of motion.
This contrasts with the remaining beams, which undergo a classical-like
motion though they do not follow the real classical trajectory; the
residual term involving $\hbar$ is a clear indicator of the nonlocal
behavior of quantum mechanics.\cite{sanz8}
Nevertheless, as $m$ increases, this deviation from the real classical
motion gets smaller and smaller but never zero (as will be shown below).

The effects derived from the noncrossing property in periodic grating
systems are equivalent to considering the presence of some effective
impenetrable barriers.
In the same way, the temporary accumulation of quantum trajectories
in certain space regions could also be thought as the effect of an
{\it effective quantum pressure}.
For instance, the particles moving along an impenetrable (effective)
barrier will keep their motion until those arriving from the central
part of the initial wave packet will move towards $x=0$, thus
decreasing the ``pressure'' exerted on the former (the same effect
has been discussed in multimode cavities\cite{sanz1} or soft double
slits\cite{sanz3}).
The decrease of the quantum pressure is responsible for the appearance
of Fraunhofer fringes; beyond the Talbot region, the particles feel a
smaller pressure that allows them to get into different channels.
After some time, particles reach the configuration equilibrium regime,
where they move along well-defined channels, the well-known Bragg
diffraction channels.


\section{Slit arrays}
\label{sec:4}

\begin{figure}
 \includegraphics[width=8cm]{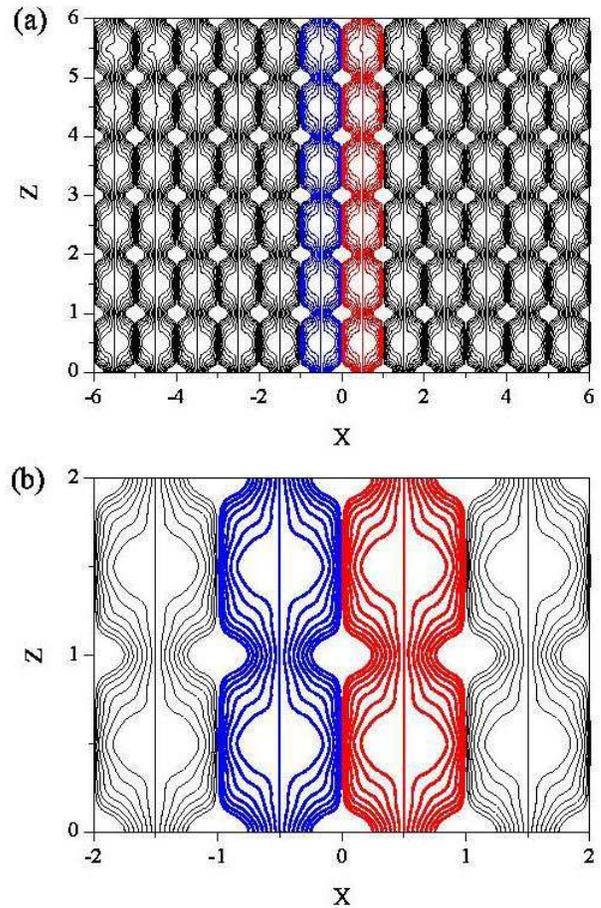}
 \caption{\label{fig3}
  (Color online) (a) Quantum trajectories associated with
  Fig.~\ref{fig1}(a).
  Trajectories in different color indicate that though according
  to Fig.~\ref{fig1}(a) the Talbot carpet seems to be a one-piece
  pattern, it is indeed constituted by many single bunches of
  trajectories arising from each slit and without crossing with
  those exiting from neighboring slits.
  (b) Enlargement of part~(a) in the region close to the slits in order
  to show the correspondence with Fig.~\ref{fig1}(b).
  In both panels, the $x$ distance is scaled in units of the grating
  period $d$, and $z$ in units of twice the Talbot distance ($2z_T$).}
\end{figure}

First we consider the dynamics due to slit arrays, where He atoms
are the diffracted particles and the slits are described by Gaussian
(transmission) functions.\cite{sanz2,sanz3}
In Fig.~\ref{fig3}(a) we can observe that, in accordance to the
previous discussion in Sec.~\ref{sec:3}, trajectories follow
the flow characterizing $\rho_t$ in Fig.~\ref{fig1}(a).
Note that here the {\it trajectory carpet} is not a one-piece
structure, as happened with the Talbot carpet generated by $\rho_t$
in Fig.~\ref{fig1}(a), but it consists of many unit substructures
(as many as slits have been considered).
These structures are the causal effect of having indistinguishable unit
cells.
Moreover, trajectories exiting from one of the slits never cross those
leaving the other slits, this being due to the noncrossing property
described above.
This can be better seen by looking at the enlargement presented in
Fig.~\ref{fig3}(b).
In this plot we observe that initially the trajectories leave the
slit in a diffusive manner towards each border of the slit.
Since all the slits are identical (as well as their transmission
function), these trajectories will feel in a short term the presence
of bunches of trajectories coming from the neighboring slits.
Then, the trajectories start bending until they move perpendicular
to the slits for a while, pushed by the neighboring trajectories.
This is a clear manifestation of the quantum pressure: the pressure
exerted by bunches of trajectories moving in opposite directions gives
rise to an effect similar to that of having an impenetrable (infinite)
potential barrier.
But the quantum pressure is also felt from the action of the
trajectories coming from the same slit: those started with initial
conditions closer to the center of the slit will push the other to
move parallel; only when they start moving back again towards the
central axis of the slit, the quantum pressure will decrease enough
as to allow the outer trajectories to move again towards the original
position of the slit.
In this way, at $z = 2z_T$ we will recover again the initial pattern
of a sum of Gaussians.
It is interesting to stress that the maxima at $z = z_T$, shown in
Fig.~\ref{fig1}(b), have not the same structure as those at $z = 0$
or $z = 2z_T$; the trajectories contributing to the maxima at $z = z_T$
start from two different (neighboring) slits, while at $z = 2z_T$ only
the trajectories started from the same slit will contribute to the
corresponding maximum.

\begin{figure}
 \includegraphics[width=8.2cm]{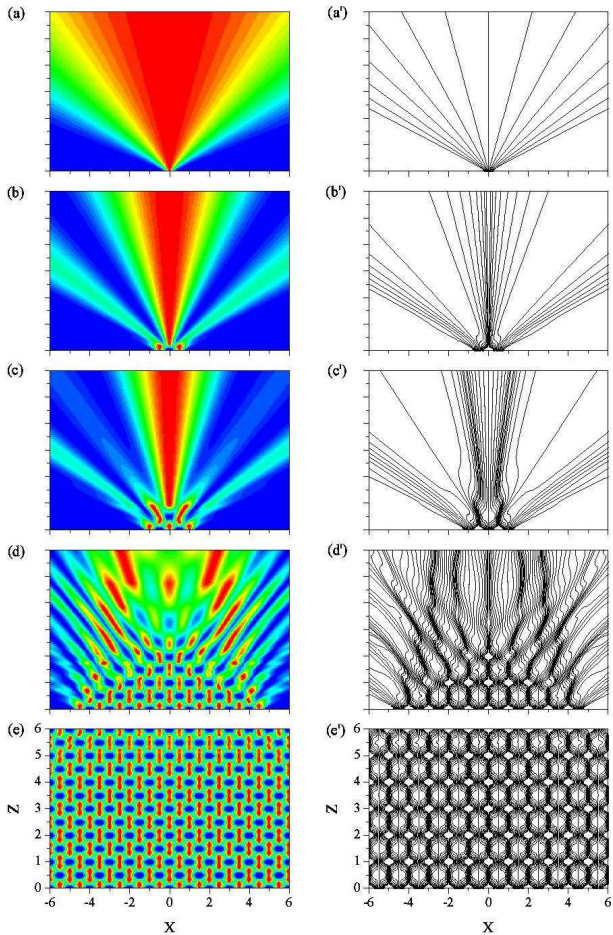}
 \caption{\label{fig4}
  (Color online) Left: Appearance of the Talbot carpet within a
  certain space region as the number of slits increases: (a) $N = 1$,
  (b) $N = 2$, (c) $N = 3$, (d) $N = 10$, and (e) $N = 50$.
  Right: Quantum trajectories corresponding to the cases shown in the
  left panels.
  In all panels, the $x$ distance is scaled in units of the grating
  period $d$, and $z$ in units of twice the Talbot distance ($2z_T$).}
\end{figure}

Provided the grating extends to infinity, the unit structures of the
pattern seen in Fig.~\ref{fig3}(a) repeat indefinitely.
However, standard slit arrays have a limited size as well
as the the incident beam.
Therefore, it is clear that the Talbot pattern will be observable only
within certain boundaries; beyond these boundaries it will start to
blur up, observing gradually the appearance of the Bragg diffraction
channels.\cite{sanz4}
Instead of going to this limit by propagating quantum trajectories
further and further away, let us rather consider the case where we
increase the number of slits progressively until reaching a number
such that within a certain bound region we have the certainty of
observing a Talbot pattern, but that further away we will only see
Fraunhofer diffraction channels.
This transition can be observed, from top to bottom, in Fig.~\ref{fig4}
with both $\rho_t$ (left) and the associated quantum trajectories
(right).

As can be seen in Figs.~\ref{fig4}(a) and \ref{fig4}(a'), the single
slit case, the lack of neighboring trajectories allows the trajectories
to spread out in all directions with no bound and therefore no (Talbot)
pattern can be observed.
Free propagation of a Gaussian wave packet just means, according to
this trajectory picture, that one can observe free motion (rectilinear
and uniform) as in classical mechanics almost since the beginning of
the propagation.
In terms of the standard quantum mechanics, this means that $\rho_t$
spreads linearly in time, but keeping its volume constant.
This constant volume only means that the number of trajectories
conserves although they are far apart from one another.

When two slits are considered, as in Figs.~\ref{fig4}(b) and
\ref{fig4}(b'), things change dramatically.
From standard quantum-mechanical viewpoint we observe a channeling
structure due to the interference of the two outgoing Gaussian wave
packets.
This is the Fraunhofer diffraction pattern.
From the quantum trajectory viewpoint, though trajectories leave along
those different channels (with a very low density of trajectories in
between), the fact that the perpendicular semiplane behind the slits
has been divided into two identical halves, where trajectories do not
cross the half one dominated by the opposite slit is remarkable.
There is a very strong quantum pressure exerted by the trajectories
arising from each half along the symmetry axis of the system.
Moreover, it is also worth commenting that very close to the slits a
certain pattern, with two temporary maxima just behind the slits, is
already present.
In Figs.~\ref{fig4}(c) and \ref{fig4}(c'), the three slit case, a
similar pattern to that described when discussing Fig.~\ref{fig3}(b)
is observed.
However, a Talbot pattern is still not formed because Fraunhofer
diffraction channels emerge immediately.

From the previous comments we can then establish that the Talbot effect
can be seen as a ``partition'' of the space as the number of slits
increases because of the strong effect (acting as an infinite barrier)
of the quantum pressure.
This is confirmed when we go to $N = 10$ [see Figs.~\ref{fig4}(d) and
\ref{fig4}(d')]  and $N = 50$ [see Figs.~\ref{fig4}(e) and
\ref{fig4}(e')].
Indeed, the existence of the quantum pressure leads to a sort of
quantum equilibrium state in which the unit structures formed by the
bunches of trajectories can coexist.
Only when the quantum pressure starts decreasing, these units begin
to blur up since the trajectories spread out the corresponding
boundaries.
Since this effect is similar to a dissipation, it could be called a
{\it quantum trajectory dissipation} (though its nature is different
to that of real dissipative phenomena).
This ``nonequilibrium'' situation remains until a new equilibrium is
established: the Fraunhofer regime.
One must realize that the stationarity of the Talbot regime
is only typical of this near field phenomenon.
In general, near field or Fresnel phenomena are not stationary.
Nonetheless, as we will see in the next section, one can still speak
about a certain class of stationarity (in the momentum space) within
the Fresnel diffraction regime different from  that observed in a
Talbot regime.


\section{Atom-surface scattering}
\label{sec:5}

\begin{figure}
 \includegraphics[width=8.2cm]{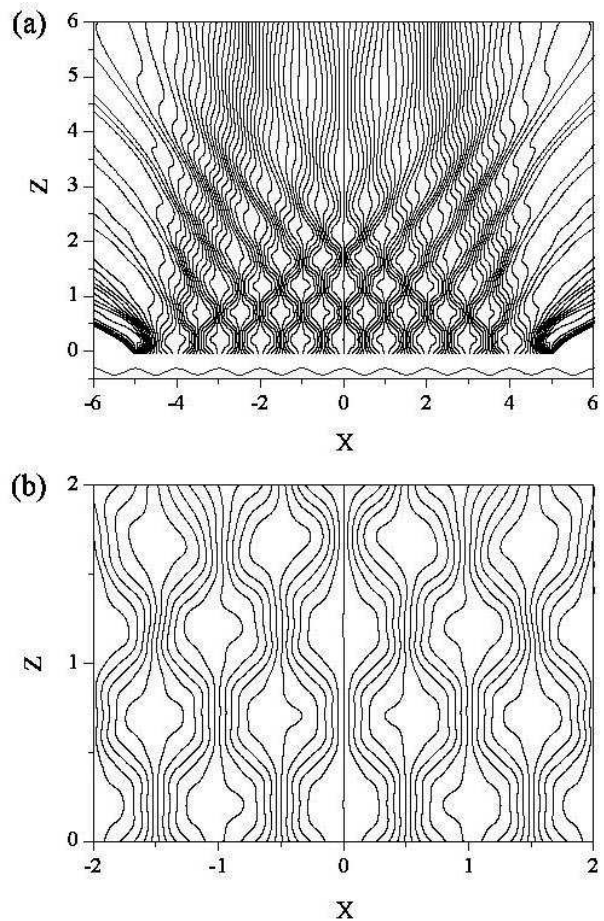}
 \caption{\label{fig5}
  (a) Quantum trajectories corresponding to the He--Cu(110) system at
  21 meV and normal incidence.
  Only the emergent part has been plotted.
  (b) Enlargement of the trajectories shown in part (a).
  In both panels, the $x$ distance is scaled in units of the Cu(110)
  unit cell $d$, and $z$ in units of twice the Talbot distance for
  an $N$-slit grating ($2z_T = 2d^2/\lambda$).}
\end{figure}

Talbot patterns appear whenever one deals with any kind of infinitely
periodic structure in which interference can be observed, and not only
with gratings consisting of slit arrays.
This is the case, for instance, of atom-surface scattering where the
surface plays the role of the periodic grating.\cite{sanz3,sanz5}
In Fig.~\ref{fig5} the formation of the Talbot pattern is plotted
when we have a beam of He atoms illuminating ten unit cells of the
Cu(110) surface at perpendicular incidence and energy $E_z = 21$~meV.
This scattering system is described by a classical interaction
potential $V(x,z) = V_M(z) + V_C(x,z)$, where
\begin{equation}
 V_{\rm M}(z) = D (1 - e^{- \alpha z})^2 - D
 \label{eq:12trsar}
\end{equation}
is a Morse potential, and
\begin{multline}
 V_{\rm C}(x,z) = D e^{-2 \alpha z}
  \left[ 0.03 \ \cos \left( \frac{2 \pi x}{d} \right) \right. \\
  + \left. 0.0004 \ \cos \left( \frac{4 \pi x}{d} \right) \right] ,
 \label{eq:13trsar}
\end{multline}
%
the coupling term between the two degrees of freedom.
Here, $D = 6.35$~meV, $\alpha = 1.05$~\AA$^{-1}$, and $d = 3.6$~\AA\
(as the aperture of the slits used before).\cite{lapu,sanz4}
The corrugation of the surface is very weak, as can be appreciated by
the amplitudes of the cosine functions.

In order to make the plot clear, we have neglected the incident part
of the trajectories and considered as before the evolution of those
corresponding to the centroid line of the incoming wave parallel to
the surface.
As can be seen in Fig.~\ref{fig5}(a), the pattern is very similar to
that displayed by the ten-slit array plotted in Fig.~\ref{fig4}(d').
This is because the corrugation of the surface is relatively weak.
This can also be seen if we compare Figs.~\ref{fig5}(b) and
\ref{fig3}(b); though the latter has been obtained considering
$N = 50$, the structure is similar in both cases (we are inside the
Talbot area).
Note, however, that the Talbot structure is slightly distorted and
repeats a bit further away
than twice the Talbot distance considered before ($2d^2/\lambda$).
The reason for this shift and distortion is the attractive
part of the interaction potential $V$, which causes a certain
acceleration in the motion of the particles.
This is the so-called {\it Beeby correction} in atom-surface
scattering.\cite{beeby}
In other words, only when the potential is flat, we can assume that
$z_T = d^2/\lambda$.
Otherwise, we should consider an effective Talbot distance,
\be
 \tilde{z}_T \equiv \frac{d^2}{\tilde{\lambda}(x,z)}
  = z_T \ \! \sqrt{1 - \frac{V(x,y)}{E_z}} ,
 \label{eqv1}
\ee
where $\tilde{\lambda}=2\pi\hbar/\sqrt{2m(E_z - V)}$.
Taking into account that $V<0$, it is clear that the square root factor
in Eq.~(\ref{eqv1}) will be greater than 1, and
therefore $\tilde{z}_T > z_T$.
The correction factor in Eq.~(\ref{eqv1}) due to the presence of the
well depth is $\sqrt{1 + D/E_z}$, and therefore we obtain
$\tilde{z}_T/z_T = 1.14$, which is basically the discrepancy observed
in Fig.~\ref{fig5}(b).
Hence, within the context of atom-surface scattering, it would be
appropriate to speak about the {\it Talbot-Beeby effect}, which gathers
both the effects caused by the periodicity (depending on $V_C$) and
those arising from the attractive part of the interaction (depending
on $V_M$).

It is clear that since there is no infinite beam of He atoms
illuminating the Cu surface, the quantum pressure will decrease as the
atoms get further away from the surface, and the Talbot pattern will
disappear.
However, there is something very interesting in this kind of systems:
one can extract already very important information about the
diffraction peaks (experimentally detected at Fraunhofer distances) once
the classical asymptotic region is reached ($V \simeq 0$).
This is a nice manifestation of the effect mentioned above: the motion
within the Fresnel region is governed by the momenta that will give
rise later on to the different Fraunhofer diffraction channels.
This has been easily proven by Sanz {\it et al.}\cite{sanz4} by
computing the $S$-matrix elements in the classical asymptotic region
and comparing them with the intensity calculated from quantum
trajectories collected in the asymptotic region.
A similar calculation for slit arrays by the same authors can be seen
in Ref.~\onlinecite{sanz2}.
The Fraunhofer regime is reached very far from the region where the
interaction potential $V$ is negligible and increases its distance
as the number of unit cells illuminated by the initial atomic beam
is increased.
Thus, for example, for ten unit cells the Fraunhofer regime is
reached around 1000~\AA.


\section{Correspondence principle and Talbot effect}
\label{sec:6}

\begin{figure}
 \includegraphics[width=8.2cm]{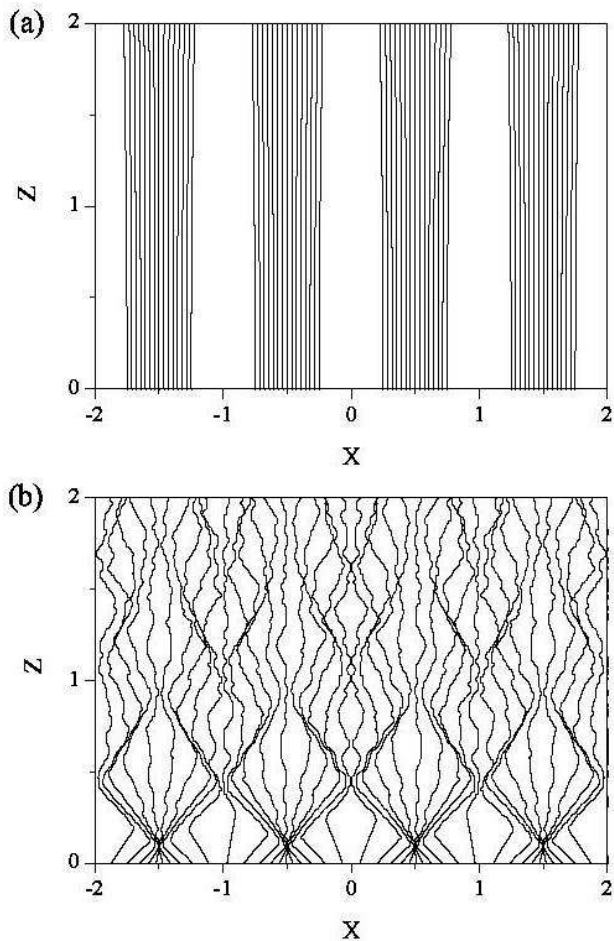}
 \caption{\label{fig6}
  Diffraction of a particle with a mass 500 times heavier than that
  of a He atom by (a) ten slits and (b) ten unit cells of the Cu(110)
  surface.
  In both panels, the $x$ distance is scaled in units of the Cu(110)
  unit cell $d$, and $z$ in units of twice the Talbot distance for
  an $N$-slit grating ($2z_T = 2d^2/\lambda$).
  Moreover, the perpendicular incidence energy is $E_z = 21$~meV in
  both cases.}
\end{figure}

As heavier impinging particles are considered, i.e., when we approach
the classical limit, the behaviors observed in previous sections
undergo dramatic changes, approaching or resembling those expected
classically.
If instead of He atoms we consider a fictitious particle with a mass
500 times that of a He atom, it is clear that quantum effects should
disappear or, at least, decrease.
This is what one could think, mistakenly, when observing
Fig.~\ref{fig6}(a), where we have represented the diffraction of
such particles by a grating consisting of ten slits.
As can be seen, the trajectories are basically straight lines, what
induces to think that no interference effects are present (neither
Talbot nor Fraunhofer ones), but particles move tracing a simple
rectilinear, uniform motion.
Obviously, this is misleading; both the $x$ and $z$ directions are
given in terms of the old $d$ and $z_T$.
However, if we replot this figure taking into account that now $z_T$
has increased by a factor $\sqrt{m/m_{\rm He}}$ (where $m$ is the mass
of the fictitious particle), we obtain again a Talbot pattern.
And far away we would find the same Fraunhofer pattern as before.
What has happened is that particles move now more slowly, and therefore
the spreading of the corresponding bunches of trajectories will also
be slower.

The previous result shows that quantum particles remain quantum even
in the so-called classical limit;\cite{sanz7} one only needs to be
patient and wait enough time in order to observe again the quantum
phenomena (of course, unless the complexity of the system is so
enormous that any quantum effect is imperceptible experimentally).
However, what happens if instead of ten slits we illuminate ten cells
of the Cu(110) surface with a beam of our fictitious particle?
The answer appears in Fig.~\ref{fig6}(b): now the topology displayed
by the quantum trajectories tries to resemble that of the classical
one, but the noncrossing principle holds.
That is, since classical trajectories give rise to the appearance
of two caustics (direction of maximum reflected intensity) at the
so-called rainbow angles, the quantum trajectories will try to describe
a similar structure, with the particularity that they cannot cross,
and therefore, after tracing an almost straight line, they will be
bounced backwards in relatively sharp angle, as seen in the lowest
part of Fig.~\ref{fig6}(b).
Moreover, notice that, because of this motion, the laminarity of the
flow described by the trajectories is lost; now the topology of the
trajectories is more irregular, this leading to some crossings (though
they occur at different times).


\section{Conclusions}
\label{sec:7}

To conclude, there are three important points in this work that are
worth stressing.
First, regardless of the gradual spreading of the partial wave function
associated with each unit cell of a periodic grating, the corresponding
initial swarms of quantum trajectories keep moving in a bound space
region.
To some extent, this is analogous to considering the trajectories
moving inside a multimode cavity.\cite{sanz1,sanz3}
Because of this effect, within the Talbot interferometry context, the
superposition principle and the Born--von Karman boundary conditions
in Bohmian mechanics have to be understood in a
very different way as they are in standard quantum mechanics.
It is not simply the overlapping of a number of wave functions but
allocation of identical copies of the same trajectory behavioral
pattern.
Second, this behavior is somehow similar to the classical behavior
observed when studying periodic structures: all the information about
the periodic structure can be obtained by simply studying the classical
effects provoked by one of the periods although nonlocal effects
are always present.\cite{sanz2}
And third, the Talbot-Beeby effect is proposed to understand quantum
trajectory carpets in the presence of attractive interaction
potentials.


\section*{Acknowledgements}

This work has been supported by the Spanish Ministry of Education and
Science under the project with reference No.~FIS2004-02461 and for a
``Juan de la Cierva'' Contract awarded to one of the authors (A.S.S.)


\end{document}